\documentclass{appolb}
\usepackage{graphicx}
\usepackage{amsmath,bm}

\DeclareMathOperator{\tr}{tr}

\begin{document}
\title{Spin 1 low lying meson spectra and the subtle link to the spin 0 mesons%
\thanks{Presented at Excited QCD 2017, Sintra. Work supported in part by Funda\c{c}\~{a}o para a Ci\^{e}ncia e a Tecnologia (FCT), through PhD grant SFRH/BD/110315/2015 and project UID/FIS/04564/2016, and Centro de F\'{i}sica da Universidade de Coimbra.} 
}
\author{J. Morais, B. Hiller
\address{CFisUC, Department of Physics, University of Coimbra, 3004-516 Coimbra, Portugal}
\\
A. A. Osipov
\address{Joint Institute for Nuclear Research, Bogoliubov Laboratory of Theoretical Physics, 141980 Dubna, Russia}
}
\maketitle
\vspace{-1cm}
\begin{abstract}
An NJL-type three-flavor quark model with a complete set of explicit chiral symmetry breaking terms is extended to include vector and axial vector
effective interaction terms. The bosonized Lagrangian is written up to quadratic order in the bosonic fields and the role of the new interactions is analysed in detail. The model's parameters are fitted to yield reasonable values to the four low-lying spin 0 and spin 1 meson nonets' masses.
\end{abstract}
\vspace{-0.25cm}

\section{Introduction - Spin 0 Model}
We discuss an effective model of QCD which is based upon the assumption that, at low-energies, the dynamical breaking of chiral symmetry plays a dominant role in the dynamics of the light mesons. The mechanism for this dynamical breaking relies on a suitable set of chirally symmetric effective multiquark vertices, in line with the ideas behind the Nambu--Jona-Lasinio (NJL) model \cite{NJL}.

Our starting point is a 3 flavor NJL-type model which includes all non-derivative effective multiquark vertices (involving products of spin 0 bilinears) which are relevant for the dynamical breaking of chiral symmetry in 4 dimensions \cite{model1,Andrianov}. These coincide with a NLO expansion in $N_c^{-1}$ (where $N_c$ is the number of colors). Besides the usual LO 4 quark term, we take into account the 6 quark 't Hooft determinant term (which is NLO in $N_c^{-1}$ and is important to break the $U\left(1\right)_A$ symmetry explicitly) and two 8 quark terms which are of the same order in $N_c$ counting as the 't Hooft term. Explicit symmetry breaking effects due to finite current quark masses are particularly important when the strange quark is involved, owing to its significantly higher mass relative to the lighter up and down quarks. The usual Dirac mass term is the LO mass dependent term in $N_c$ counting, but consistency with the order to which chirally symmetric terms are included demands that we include all explicit breaking terms to NLO as well. By letting the quark fields interact with an external scalar source field, we may systematically enumerate all such explicit symmetry breaking NLO terms.

We define the quark bilinears $s_a = \bar{q} \lambda_a q$, $p_a = \bar{q} i \gamma_5 \lambda_a q$, the $U\left(3\right)$ valued field $\Sigma = \frac{1}{2} \left(s_a - i p_a\right)\lambda_a$, and an external source $\chi$ which is assumed to transform as $\Sigma$. Since the model is non-renormalizable, we scale each effective vertex by an appropriate power of $\Lambda$, which is supposed to be of the order of the chiral symmetry breaking scale $\Lambda_{\chi SB} \sim 1 \, \text{GeV}$. The chirally symmetric terms read
\begin{align}
\label{main0}
\mathcal{L}_{int} & = \frac{\bar{G}}{\Lambda^2} \tr{\left(\Sigma^\dagger \Sigma\right)} + \frac{\bar{\kappa}}{\Lambda^5} \left(\det{\Sigma} + \det{\Sigma^\dagger}\right) \nonumber \\
& + \frac{\bar{g}_1}{\Lambda^8} \left(\tr{\Sigma^\dagger \Sigma}\right)^2 + \frac{\bar{g}_2}{\Lambda^8} \tr{\left(\Sigma^\dagger \Sigma \Sigma^\dagger \Sigma\right)}.
\end{align}

\noindent Explicit symmetry breaking terms are constructed by suitable replacements $\Sigma \to \chi$ in $\mathcal{L}_{int}$ terms. This process yields the following additional terms:
\begin{align}
\label{0esb}
\mathcal{L}_0 & = -\tr{\left(\Sigma^\dagger \chi + \chi^\dagger \Sigma\right)}, & \mathcal{L}_5 & = \frac{\bar{g}_5}{\Lambda^4} \tr{\left(\Sigma^\dagger \chi \Sigma^\dagger \chi\right)} + h.c., \nonumber \\
\mathcal{L}_2 & = \frac{\bar{\kappa}_2}{\Lambda^3} \epsilon_{ijk} \epsilon_{mnl} \Sigma_{im} \Sigma_{jn} \chi_{kl} + h.c., & \mathcal{L}_6 & = \frac{\bar{g}_6}{\Lambda^4} \tr{\left(\Sigma^\dagger \Sigma \chi^\dagger \chi\right)} + h.c., \nonumber \\
\mathcal{L}_3 & = \frac{\bar{g}_3}{\Lambda^6} \tr{\left(\Sigma^\dagger \Sigma \Sigma^\dagger \chi\right)} + h.c., & \mathcal{L}_7 & = \frac{\bar{g}_7}{\Lambda^4} \left(\tr{\Sigma^\dagger \chi} + h.c.\right)^2, \nonumber \\
\mathcal{L}_4 & = \frac{\bar{g}_4}{\Lambda^6} \tr{\left(\Sigma^\dagger \Sigma\right)} \tr{\left(\Sigma^\dagger \chi\right)} + h.c., & \mathcal{L}_8 & = \frac{\bar{g}_8}{\Lambda^4} \left(\tr{\Sigma^\dagger \chi} - h.c.\right)^2 .
\end{align} 

\noindent Substituting $\chi \to \frac{m}{2}$ allows us to identify $\mathcal{L}_0$ with the usual Dirac mass term, which together with the 4 quark term in $\mathcal{L}_{int}$ form the LO effective contributions in $N_c$ counting. All other terms constitute a complete set of NLO terms describing both dynamical and explicit breaking of chiral symmetry (there would be 3 additional terms related to the known Kaplan-Manohar ambiguity which have been set to 0 without loss of generality).

The terms proportional to $\kappa,\kappa_1,\kappa_2,g_1,g_4,g_7,g_8,g_{10}$ trace OZI rule violating affects, while those proportional to $g_2,g_3,g_5,g_6,g_9$ express a contribution of four-quark components $\bar{q}q\bar{q}q$ to the quark content of the mesons. 

The model was fitted to successfully reproduce the low-lying scalar and pseudoscalar meson spectra, as well as a number of strong and radiative meson decays. It was employed to study the QCD phase diagram and equation of state, and to assess the possibility of strange quark matter \cite{SQM}.

\section{Inclusion of Spin 1 Mesons}

Following the same methodology that has been used in the spin 0 version of the model, we define the quark bilinears $v_a^{\mu} = \bar{q} \gamma^{\mu}\lambda_a q$,  $a_a^{\mu} = \bar{q} \gamma^{\mu} \gamma_5 \lambda_a q$, and the $U\left(3\right)$ valued fields $R^{\mu},L^{\mu} = \frac{1}{2} \left(v_a^{\mu} \pm a_a^{\mu}\right)\lambda_a$. From these and previous definitions, we extend the full set of effective vertices 
to \cite{model2}:
\begin{align}
\mathcal{L}'_1 & = \frac{\bar{w}_1}{\Lambda^2} \tr{\left(R^{\mu}R_{\mu} + L^{\mu}L_{\mu}\right)}, & \mathcal{L}'_5 & = \frac{\bar{w}_5}{\Lambda^8} \tr{\left(R^{\mu}R_{\mu}R^{\nu}R_{\nu} + L^{\mu}L_{\mu}L^{\nu}L_{\nu}\right)}, \nonumber \\
\mathcal{L}'_2 & = \frac{\bar{w}_2}{\Lambda^8} \left[\tr{\left(R^{\mu}R_{\mu} + L^{\mu}L_{\mu}\right)}\right]^2, & \mathcal{L}'_6 & = \frac{\bar{w}_6}{\Lambda^8} \tr{\left(R^{\mu}R_{\mu} + L^{\mu}L_{\mu}\right)} \tr{\left(\Sigma^\dagger \Sigma\right)}, \nonumber \\
\mathcal{L}'_3 & = \frac{\bar{w}_3}{\Lambda^8} \left[\tr{\left(R^{\mu}R_{\mu} - L^{\mu}L_{\mu}\right)}\right]^2, & \mathcal{L}'_7 & = \frac{\bar{w}_7}{\Lambda^8} \tr{\left(\Sigma^\dagger L^{\mu} \Sigma R_{\mu}\right)}, \nonumber \\
\mathcal{L}'_4 & = \frac{\bar{w}_4}{\Lambda^8} \tr{\left(R^{\mu}R^{\nu}R_{\mu}R_{\nu} + L^{\mu}L^{\nu}L_{\mu}L_{\nu}\right)}, & \mathcal{L}'_8 & = \frac{\bar{w}_8}{\Lambda^8} \tr{\left(\Sigma^\dagger \Sigma R^{\mu} R_{\mu} + \Sigma \Sigma^\dagger L^{\mu} L_{\mu}\right)}, \nonumber
\end{align}
\begin{align}
\mathcal{L}'_9 & = \frac{\bar{w}_9}{\Lambda^6} \tr{\left(R^{\mu}R_{\mu} + L^{\mu}L_{\mu}\right)} \tr{\left(\Sigma^\dagger \chi + \Sigma \chi^\dagger\right)}, & \mathcal{L}'_{12} & = \frac{\bar{w}_{12}}{\Lambda^4} \tr{\left(\chi^\dagger L^{\mu} \chi R_{\mu}\right)}, \nonumber \\
\mathcal{L}'_{10} & = \frac{\bar{w}_{10}}{\Lambda^6} \tr{\left(\chi^\dagger L^{\mu} \Sigma R_{\mu} + \Sigma^\dagger L^{\mu} \chi R_{\mu}\right)}, & \mathcal{L}'_{13} & = \frac{\bar{w}_{13}}{\Lambda^4} \tr{\left(\chi^\dagger \chi R^{\mu} R_{\mu} + \chi \chi^\dagger L^{\mu} L_{\mu}\right)}, \nonumber
\end{align}
\begin{equation}
\mathcal{L}'_{11} = \frac{\bar{w}_{11}}{\Lambda^6} \tr{\left[\left(\Sigma^\dagger \chi + \chi^\dagger \Sigma\right) R^{\mu} R_{\mu} + \left(\Sigma \chi^\dagger + \chi \Sigma^\dagger\right) L^{\mu} L_{\mu} \right]}.
\end{equation}
\noindent This extension introduces 13 new parameters to the model, but it can be shown that only a subset contributes to the vacuum properties of the model.

As has been done with the spin 0 version of the model, the model is bosonized in a functional integral formalism. In order to describe the system in the broken Nambu-Goldstone phase, we include a shift $\sigma \to \sigma + M$ and interpret $M$ as a constituent quark mass matrix. The bosonization procedure involves rewriting multiquark interactions in terms of the auxiliary $s_a$, $p_a$, $v_a^{\mu}$, and $a_a^{\mu}$ fields. This static part of the functional integral is evaluated in a stationary phase approximation using the following expansions:
\begin{align}
\label{SPA-exp}
s_a^{st} & = h_a + h_{ab}^{\left(1\right)} \sigma_b + h_{abc}^{\left(1\right)} \sigma_b \sigma_c + h_{abc}^{\left(2\right)} \phi_b \phi_c + H_{abc}^{\left(1\right)} V^{\mu}_b V_{c\mu} + H_{abc}^{\left(2\right)} A^{\mu}_b A_{c\mu} + \dots \nonumber \\
p_a^{st} & = h_{ab}^{\left(2\right)} \phi_b + h_{abc}^{\left(3\right)} \phi_b \sigma_c + H_{abc}^{\left(3\right)} V^{\mu}_b A_{c\mu} + \dots \nonumber \\
v^{\mu\, st}_a & = H_{ab}^{\left(1\right)} V^{\mu}_b + H_{abc}^{\left(4\right)} \sigma_b V^{\mu}_c + H_{abc}^{\left(5\right)} \phi_b A^{\mu}_c + \dots \nonumber \\
a^{\mu\, st}_a & = H_{ab}^{\left(2\right)} A^{\mu}_b + H_{abc}^{\left(6\right)} \phi_b V^{\mu}_c + H_{abc}^{\left(7\right)} \sigma_b A^{\mu}_c + \dots
\end{align}

\noindent Here, $h_a$ may be identified with the quark condensates, while the other $h$ and $H$ coefficients provide effective contributions to the masses and couplings of the various meson fields. All coefficients are expressed recursively in terms of lower order ones. The remaining quark determinant is evaluated in a generalized heat kernel expansion approach.

The quadratic part of the bosonized Lagrangian includes mixing terms 
\begin{equation}
\label{mixing}
\tr{\left(i \left[V^{\mu},M\right]\partial_{\mu}\sigma - \left\lbrace A^{\mu},M \right\rbrace \partial_{\mu}\phi \right)} ,
\end{equation}

\noindent which require the following shift definitions \cite{model2,diag} in order to be eliminated:
\begin{align}
\label{shifts}
&V_{a\mu}\rightarrow V_{a\mu}+k_a X_{a\mu}, \quad A_{a\mu}\rightarrow A_{a\mu}+k'_a Y_{a\mu}, \\
&X_{a\mu}=2f_{abc}M_b\partial_{\mu}\sigma_c, \quad Y_{a\mu}=2d_{abc}M_b\partial_{\mu}\phi_c.
\end{align}

\noindent Due to 8 quark interactions and explicit symmetry breaking, mass diagonalization conditions lead to generally different $k_a$ and $k'_a$. These shifts contribute to spin 0 mesons' kinetic terms, leading to new (unsymmetric) renormalizations for these fields. Spin 1 meson fields are still renormalized in the usual way. In the isospin approximation ($m_u = m_d \neq m_s$), mixing angles between neutral strange and non-strange components in the scalar and pseudoscalar sectors must still be taken care of through the introduction of mixing angles $\psi_{\sigma}$, $\psi_{\phi}$.

The masses of the spin 1 mesons are obtained from the bosonized Lagrangian as
\begin{align}
M_{\rho}^2 & =M_{\omega}^2=\frac{3}{2}\varrho^2 H_{11}^{\left(1\right)}, & M_{a_1}^2 & = M_{f_1}^2=\frac{3}{2}\varrho^2H_{11}^{\left(2\right)}+6 M_{u}^2 \nonumber \\ 
M_{K^*}^2 & =\frac{3}{2}\left[\varrho^2 H_{44}^{\left(1\right)}+\left(M_{u}-M_s\right)^2 \right], & M_{K_1}^2 & = \frac{3}{2}\left[\varrho^2 H_{44}^{\left(2\right)}+\left(M_{u}+M_s\right)^2\right] \nonumber \\ 
M_{\varphi}^2 & = 3\varrho^2H_{ss}^{\left(1\right)} & M_{f'_1}^2 & = 3 \varrho^2H_{ss}^{\left(2\right)}+6 M_s^2 ,
\end{align}

\noindent where $\varrho$ is a quantity related with the quark loop integrals arising within the heat kernel approach. The $H$ coefficients depend only on the values of the new $w_i$ parameters, which means that no direct information from the spin 0 sector contributes to spin 1 mesons' masses. In turn, owing to the form of the mixing (\ref{mixing}), spin 1 related coefficients $H$ will appear in the expressions for the masses of the spin 0 mesons. This results in direct relations between the squared masses of spin 0 and spin 1 mesons. For example, we have that
\begin{align}
M_{a_0}^2 & = \frac{2}{3 H_{11}^{\left(1\right)}} \left(\frac{h_{u}}{M_{u}}- h_{11}^{\left(1\right)}\right) M_{\rho}^2 + 4 M_{u}^2 \nonumber \\
M_{\eta}^2 & = \frac{1}{1-\tan^2{\psi_{\phi}}} \frac{1}{3H_{uu}^{\left(2\right)}} \left(\frac{h_{u}}{M_{u}} - 2 h_{uu}^{\left(2\right)} - 2 h_{ud}^{\left(2\right)}\right) M_{f_1}^2 \nonumber \\
& + \frac{1}{1-\cot^2{\psi_{\phi}}} \frac{1}{3H_{ss}^{\left(2\right)}} \left(\frac{h_s}{M_s} - 2 h_{ss}^{\left(2\right)}\right) M_{f'_1}^2 ,
\end{align}

\noindent Essentially, there is a full set of linear expressions of the form $M_{s,p}^2 = c_1 M_{v,a}^2 + c_2$, one for each homologous pair of spin 0 and spin 1 mesons, with slightly modified forms for the mixed neutral channels which depend on $\psi_{\sigma,\phi}$. These relations reflect the strict symmetry constraints intrinsically built into the model.

\section{Parameter Fitting}

Of the 13 new parameters, only 9 appear in the $H$ coefficients which are relevant for the mass spectra ($w_2$ to $w_5$ do not contribute to the vacuum properties of the model). Among those, there are 3 ($w_7$,$w_{10}$,$w_{12}$) which appear with opposite sign in $H^{\left(1,2\right)}$ coefficients, leading to relations which constrain them tightly according to mass differences between spin 1 chiral partner mesons. An interesting set of 3 relations that can be derived is
\begin{align}
2\left(M_{K^*}^2 - \frac{3}{2} \left(M_u - M_s\right)^2\right)^{-1} + 2\left(M_{K_1}^2 - \frac{3}{2} \left(M_u + M_s\right)^2\right)^{-1} = \nonumber \\
M_{\rho}^{-2} + M_{\varphi}^{-2} + \left(M_{a_1}^2 - 6 M_u^2\right)^{-1} + \left(M_{f_1}^2 - 6 M_s^2\right)^{-1} ,\nonumber\\
M_{a_1}^2=\frac{6 M_{u}^4}{M_{u}^2-\varrho^2f_{\pi}^2}, \qquad
M_{K_1}^2=\frac{\frac{3}{2}\left(M_{u}+M_s\right)^4}{\left(M_{u}+M_s\right)^2-4\varrho^2f_K^2},
\end{align}

\noindent These provide a way do directly determine $M_u$, $M_s$ and $\Lambda$ (which appears in $\varrho$) from the empirical spin 1 masses together with $f_{\pi}$ and $f_K$, without any reference to neither the model's parameters nor the spin 0 spectra.

\begin{table}
	\caption{Empirical input used to fit the model's parameters (values in MeV).}
	\label{emp_input}
	\centering
	\begin{tabular}{cccccccccc}
		\hline
		\textbf{$M_\pi$} 
		&\textbf{$M_K$} 
		&\textbf{$M_\eta$}
		&\textbf{$M_{\eta'}$}
		&\textbf{$M_\sigma$}
		&\textbf{$M_\kappa$}
		&\textbf{$M_{a_0}$}
		&\textbf{$M_{f_0}$}
		&\textbf{$M_\rho$}
		&\textbf{$M_{K^*}$}
		\\ 
		\hline
		138
		& 496 
		& 548 
		& 958 
		& 500 
		& 850 
		& 980 
		& 980 
		& 778 
		& 893 \\
		\hline
		\hline
		\textbf{$M_\varphi$}
		&\textbf{$M_{a_1}$}
		&\textbf{$M_{K_1}$}
		&\textbf{$M_{f_1}$}
		&\textbf{$m_u$}
		&\textbf{$m_s$}
		&\textbf{$f_\pi$}
		&\textbf{$f_K$}
		&\textbf{$\theta_\phi$} & \\
		\hline
		1019
		& 1270 
		& 1274 
		& 1426 
		& 4 
		& 100 
		& 92 
		& 111 
		& -15$^{\circ}$ &
		\\
		\hline
	\end{tabular}
\end{table}

\begin{table}
	\caption{Results of our fit (values of $M_i$, $\Lambda$ in MeV).}
	\label{fit_results}
	\centering
	\begin{tabular}{cccccccc}
		\hline
		  \textbf{$\theta_\sigma$}
		& \textbf{$\Lambda$} 
		& \textbf{$M_u$}
		& \textbf{$M_s$}
		& \textbf{$\bm{w_1}$} 
		& \textbf{$\bm{w_6}$} 
		& \textbf{$\bm{w_9}$}
		& \textbf{$\bm{w_{13}}$} 
		\\
		\hline
		  25.1$^{\circ}$
		& 1633
		& 244
		& 508
		& -10 
		& 0
		& 0
		& 0\\
		\hline
	\end{tabular}
\end{table}

\begin{table}
	\caption{Values of non-zero parameters in natural units.}
	\label{natural}
	\centering
	\begin{tabular}{ccccccccc}
		\hline
		\text{c}
		& \textbf{$G$} 
		& \textbf{$\kappa$} 
		& \textbf{$g_1$} 
		& \textbf{$g_2$} 
		& \textbf{$\kappa_2$} 
		& \textbf{$g_3$} 
		& \textbf{$g_4$} 
		& \textbf{$g_5$} \\ 
		\hline
		\text{S}
		& $\frac{f^2\Lambda^2}{M^2}$
		& $\frac{f^4\Lambda^4}{M^3}$
		& $\frac{f^6\Lambda^6}{M^4}$
		& $\frac{f^6\Lambda^6}{M^4}$
		& $\frac{f^2\Lambda^2}{M}$
		& $\frac{f^4\Lambda^4}{M^2}$
		& $\frac{f^4\Lambda^4}{M^2}$
		& $f^2\Lambda^2$ \\
		\text{$\bar c$}
		&  1.0
		& -0.1 
		&  0.05
		& -0.1 
		&  0.01 
		& -1.3 
		&  0.3 
		& -0.5 \\ 
		\hline
		\hline
		\textbf{$g_6$}
		& \textbf{$g_7$} 
		& \textbf{$g_8$} 
		& \textbf{$w_1$} 
		& \textbf{$w_7$} 
		& \textbf{$w_8$} 
		& \textbf{$w_{10}$} 
		& \textbf{$w_{11}$} 
		& \textbf{$w_{12}$} \\
		\hline
		$f^2\Lambda^2$
		& $f^2\Lambda^2$
		& $f^2\Lambda^2$ 
		& $f^2$ 
		& $\frac{f^6\Lambda^4}{M^2}$
		& $\frac{f^6\Lambda^4}{M^2}$
		& $f^4\Lambda^2$
		& $f^4\Lambda^2$
		& $f^2M^2$ \\
		-2.6
		& -0.7 
		& -0.5 
		& -0.1 
		& -0.1 
		&  0.1 
		& -0.5 
		&  0.3 
		& -0.8 \\
		\hline
		
	\end{tabular}
\end{table}

In table \ref{emp_input} we show the empirical input used to fit the model's parameters, and in table \ref{fit_results} we show the fit results. Due to the way the $w_i$ enter in the $H$ coefficients, they are not all independent; e.g., $w_1$, $w_6$ and $w_9$ combine effectively as a single parameter. This means that the model has a high degree of degeneracy among parameter sets which yield the same vacuum results, but this is expected to be lifted upon introducing thermodynamic parameters. The effective couplings in table \ref{fit_results} have been externally fixed to the values shown there, and the rest have been fitted to the values reported in table \ref{natural}. These are shown in natural units ($\bar{c} = S c$) after a methodical removal of the various relevant scales \cite{Manohar,model2}, including not only $\Lambda$ (which estimates the scale of spontaneous symmetry breaking), but also the constituent mass $M$ (which is characteristic of chirality violations at the quark vertices) and the weak decay constant $f$ (which governs the dynamics of the pseudo-Goldstone bosons).

The fitted values of $\theta_{\sigma}$, $\Lambda$, and $M_i$ are all very reasonable. The inclusion of spin 1 mesons yields a value of $\Lambda$ which is roughly double of what it was in the spin 0 version of the model, but still well within $\mathcal{O}\left(1 \text{GeV}\right)$, and also enhances the difference $M_s - M_u$. 

Overall, we have been able to fit the model to reproduce the full 4 low-lying meson nonets' spectra. Also, we have clearly verified that the model's parameters are severely constrained by symmetry requirements. A particularly interesting manifestation of these constraints is in the subtle relations which arise between spin 0 and spin 1 masses due to the specific form of the $V-\sigma$ and $A-\phi$ mixing terms.


\begin{thebibliography}{99}
\bibitem{NJL} Y.  Nambu, G.  Jona-Lasinio,  Phys. Rev. 122,  345 (1961); Phys. Rev. 124, 246 (1961).
\bibitem{model1} A. A. Osipov, B. Hiller, A. H. Blin,  Eur. Phys. J. A49, 14 (2013); A. A. Osipov, B. Hiller, A. H. Blin, Phys. Rev. D88, 054032 (2013).
\bibitem{Andrianov} A. A. Andrianov, V. A. Andrianov, Int. J. Mod. Phys. A 08, 1981 (1993)
\bibitem{SQM} J. Moreira, J. Morais, B. Hiller, A. A. Osipov, A. H. Blin, Phys. Rev. D91, 116003 (2015).
\bibitem{model2} J. Morais, B. Hiller, A. A. Osipov, Phys. Rev. D 95, 074033 (2017).
\bibitem{diag} J. Morais, B. Hiller, A. A. Osipov, e-Print: arXiv:1705.04644 [hep-ph].
\bibitem{Manohar} A. Manohar, H. Georgi, Nucl. Phys. B234, 189 (1984).
%
%
\end{thebibliography}

\end{document}